\documentclass[journal]{IEEEtran}
\usepackage{graphicx,amssymb,amsmath}
\usepackage{multicol}
\usepackage[noadjust]{cite}
\usepackage{setspace}
\usepackage{stfloats}
\usepackage{midfloat}
\usepackage[normal]{threeparttable}
\usepackage{flushend,cuted}
\usepackage{cuted}
\usepackage{cases,subeqnarray}
\usepackage{bm,multirow,bigstrut}
\usepackage{textcomp}
\usepackage{latexsym,bm}
\usepackage{booktabs,changebar}
\usepackage{xcolor}
\usepackage{mathtools}

\IEEEoverridecommandlockouts
\begin{document}
\title{On the Ergodic Capacity of MIMO Free-Space Optical Systems over Turbulence Channels}

\author{Jiayi~Zhang,~\IEEEmembership{Member,~IEEE},
        Linglong~Dai,~\IEEEmembership{Senior Member,~IEEE},
        Yanjun~Han,~\IEEEmembership{Student Member,~IEEE},
        Yu~Zhang,~\IEEEmembership{Senior Member,~IEEE},
        and~Zhaocheng~Wang,~\IEEEmembership{Senior Member,~IEEE}
\thanks{
Manuscript received May 28, 2014; revised November 09, 2014 and March 19, 2015; accepted
April 10, 2015. This work was supported by the National Key Basic Research Program of China (No. 2013CB329203), National Natural Science Foundation of China (Grant No. 61201185), National High Technology Research and Development Program of China (Grant No. 2014AA01A704), and China Postdoctoral Science Foundation (No. 2014M560081).}%
\thanks{The authors are with Department of Electronic Engineering as well as Tsinghua
National Laboratory of Information Science and Technology (TNList), Tsinghua University, Beijing 100084, P. R. China (e-mails: \{jiayizhang, daill\}@tsinghua.edu.cn, hanyj11@mails.tsinghua.edu.cn, \{zhang-yu, zcwang\}@tsinghua.edu.cn).}}

\maketitle

\begin{abstract}
The free-space optical (FSO) communications can achieve high capacity with huge unlicensed optical spectrum and low operational costs. The corresponding performance analysis of FSO systems over turbulence channels is very limited, especially when using multiple apertures at both transmitter and receiver sides. This paper aim to provide the ergodic capacity characterization of multiple-input multiple-output (MIMO) FSO systems over atmospheric turbulence-induced fading channels. The fluctuations of the irradiance of optical channels distorted by atmospheric conditions is usually described by a gamma-gamma ($\Gamma \Gamma$) distribution, and the distribution of the sum of $\Gamma \Gamma$ random variables (RVs) is required to model the MIMO optical links. {We use an $\alpha$-$\mu$ distribution to efficiently approximate the probability density function (PDF) of the sum of independent and identical distributed $\Gamma\Gamma$ RVs through moment-based estimators.} Furthermore, the PDF of the sum of independent, but not necessarily identically distributed $\Gamma \Gamma$ RVs can be efficiently approximated by a finite weighted sum of PDFs of $\Gamma \Gamma$ distributions. Based on these reliable approximations, novel and precise analytical expressions for the ergodic capacity of MIMO FSO systems are derived. Additionally, we deduce the asymptotic simple expressions in high signal-to-noise ratio regimes, which provide useful insights into the impact of the system parameters on the ergodic capacity. Finally, our proposed results are validated via Monte-Carlo simulations.
\end{abstract}


\IEEEpeerreviewmaketitle
\section{Introduction}
Free-space optical (FSO) communication is one of the most promising technologies for indoor and outdoor wireless applications due to its merits of free license, effective cost and high bandwidth. The research of FSO systems has been taken around for the last three decades and is currently attracting more attention as the demand for high data rate continues to increase \cite{jovicic2013visible,xu2008ultraviolet}. The typical applications of FSO systems include ``last mile" access, indoor positioning, disaster recovery, military applications, underwater system, device-to-device communications, and video transmission, etc. \cite{simpson2012smart,ghassemlooy2012optical,kedar2004urban,chatzidiamantis2011inverse}.

However, the FSO communication still faces many challenges. One of the main challenges arises from the atmospheric turbulence, which is induced by the fluctuations in the atmosphere because of inhomogeneities in pressure and temperature changes, especially for over link distances of 1 km and above \cite{ghassemlooy2012optical}. The performance of FSO systems is highly vulnerable to the atmospheric channel turbulence between the transmitter and the receiver, and the parameters of the channel such as the length, the optical wavelength, and turbulence. Therefore, accurate modeling for the distribution of the turbulence-induced fading is significant to assess the performance of FSO communications. To describe weak or strong atmospheric turbulence fading, the gamma-gamma ($\Gamma \Gamma$) distribution has been proved to provide good agreement between theoretical and experimental data \cite{andrews2005laser,wang2010moment,al2001mathematical}.

The performance of FSO systems over $\Gamma \Gamma$ fading channels has been widely investigated in the literature \cite{nistazakis2009average,gappmair2011further,uysal2004error,uysal2006error,gappmair2009error}. These research have proved that effective fading-mitigation techniques are required to satisfy the typical bit error rate (BER) and capacity targets for FSO applications at the range of practical signal-to-noise ratios (SNRs). One of the promising approaches to mitigate the degrading effects of atmospheric turbulence is the use of multiple-input multiple-output (MIMO) technique, which has been widely used in wireless radio systems. By deploying multiple apertures at the transmitter and/or the receiver, the FSO system performance can be significantly enhanced. Therefore, many researchers have focused on the performance analysis of MIMO FSO systems over turbulence channels \cite{navidpour2007ber,bayaki2009performance,tsiftsis2009optical,letzepis2009outage,peppas2011multivariate,farid2012diversity}. In \cite{navidpour2007ber,bayaki2009performance,tsiftsis2009optical}, the authors have studied the BER, diversity gain and combining gain of MIMO FSO systems with equal gain combining (EGC) or maximal ratio combining (MRC) at the receiver. The outage probability of FSO communication systems over $\Gamma \Gamma$ fading channels with spatial diversity has been provided in \cite{letzepis2009outage,peppas2011multivariate,farid2012diversity}.

The performance of MIMO FSO systems over $\Gamma \Gamma$ fading channels has been extensively investigated in terms of outage probability and error rate, however, there have been few studies on the theoretical ergodic capacity. This is due to the fact that the integral for ergodic capacity is mathematically difficult to be solved. {Only very recently, the ergodic capacity of MIMO FSO communications using multiple partially coherent beams propagation through strong turbulence channels has been investigated in \cite{deng2013capacity} by using a single $\Gamma \Gamma$ approximation. However, the approximation results presented in \cite{deng2013capacity} need an adjustment parameter to obtain a sufficient approximation accuracy. In this paper, we use a more accurate approximation method to evaluate the capacity of MIMO FSO systems.}

Capitalizing on the aforementioned observations, in this work we provide the cumbersome statistical ergodic capacity analysis of MIMO FSO systems with EGC over turbulence channels, which are modeled as independent and identically distributed (i.i.d.) and independent, but not necessarily identically distributed (i.n.i.d.) $\Gamma \Gamma$ distributions, respectively. In particular, the contributions of this paper are summarized as:
\begin{itemize}
\item {Accurate analytical expressions for the ergodic capacity of MIMO FSO systems with EGC over i.i.d. $\Gamma \Gamma$ fading channels have been derived.} These expressions are given in terms of either Meijer's $G$- or Fox's $H$-functions. The Meijer's $G$-function can be easily evaluated and efficiently programmed in most standard software packages (e.g., MAPLE, MATHEMATICA), while numerically efficient methods can be used to evaluate the Fox's $H$-function \cite{peppas2012ol,yilmaz2009product}. The presented technique of ergodic capacity analysis is founded on approximating the probability density function (PDF) of the SNR at the receiver by using the $\alpha$-$\mu$ distribution \cite{yacoub2007alpha}. Compared with the approach employing in \cite{deng2013capacity}, the $\alpha$-$\mu$ approximation of the sum of multiple $\Gamma \Gamma$ random variables (RVs) is more accurate and avoid a correcting factor to obtain a sufficient approximation accuracy.

\item Based on the nested finite weighted PDF of the sum of i.n.i.d. $\Gamma \Gamma$ RVs presented in \cite{chatzidiamantis2011distribution}, we also derive a novel and analytical expression for the ergodic capacity of MIMO FSO systems for the case of i.n.i.d. $\Gamma \Gamma$ fading scenario.

\item To further provide the useful yet simple insights into the impact of system and channel parameters on the ergodic capacity, we present the asymptotic ergodic capacity bounds of MIMO FSO systems over i.i.d. and i.n.i.d. $\Gamma \Gamma$ turbulence channels in the high-SNR regime. For example, it shows that the ergodic capacity increases as the number and diameter of apertures gets larger. Interestingly, it is not a proportional relationship between the ergodic capacity and link distance as expected.
\end{itemize}

The remainder of the paper is structured as follows: Section \ref{se:2} describes the system model and the $\Gamma \Gamma$ distribution of the turbulence fading. In Section \ref{se:MIMO}, we present the approximated PDFs of the sum of multiple i.i.d. and i.n.i.d. $\Gamma \Gamma$ RVs, and derive exact and high-SNR approximation expressions for the ergodic capacity. Numerical and Monte-Carlo simulation results are provided in Section \ref{se:numerical_results}, and Section \ref{se:conclusion} concludes this paper with a summary of the main results.

\emph{Notation}: We use upper and lower case boldface to denote matrices and vectors. The expectation and variance is given by ${\text E}\left(\cdot\right)$ and ${\text {Var}}\left(\cdot\right)$, respectively. The real part of a number is given by $\Re\left(\cdot\right)$. The matrix determinant is given by $\text {det}(\cdot)$, while the Hermitian operation of a matrix $\pmb{X}$ is defined as ${\pmb{X}}^H$. The set of positive integer numbers is expressed via ${\mathbb{Z}}^+$. Finally, let $ {X}_{ij}$ denote the $\left(i,j\right)$-th element of the matrix $\pmb{X}$.

\section{FSO System Model}\label{se:2}
{We consider a typical outdoor MIMO FSO system with $M$ transmit lasers and $N$ receive apertures with same hardware over atmospheric turbulence channels.}
Similar to the related literature \cite{bayaki2009performance,chatzidiamantis2011distribution,wilson2005optical,peppas2013capacity}, {we assume that the high-energy FSO system adopts the commonly used intensity modulation/direct detection(IM/DD) of on-off keying (OOK) signals using interleaving coding.} {In this case, the MIMO FSO system's performance is limited by background radiation and thermal noise, which can be modeled as i.i.d. additive white Gaussian noise (AWGN) as a accurate approximation of the Poisson photon-counting detection model {\cite{navidpour2007ber,bayaki2009performance,peppas2013capacity,deng2013capacity}}}. Then, the received signal at the $n$th receive aperture can be expressed as
\begin{align}\label{eq:MIMO}
 {y_n = \sqrt{P_\text{avg}}\eta s I_n + v_n,}
\end{align}
where $s\in \{0, 1\}$ denotes the transmitted signal, {$P_\text{avg}$ is the transmitted average power}, $\eta$ represents the optical-to-electrical conversion coefficient, and $v_n$ is the AWGN with zero mean and variance of $\sigma_v^2=N_0/2$. Moreover, ${I_n} = \sum\nolimits_{m = 1}^M {{I_{mn}}} $ denotes the sum of $M$ fading gains at the $n$th receive aperture, where $I_{mn}$ is the irradiance from the $m$th transmit laser to the $n$th receive aperture. Since the typical FSO system provides high data rate (about $10^9$ bits/s), the time scales of these fading processes $I_{mn}$ are far larger than the bit interval (about $10^{-9}$ s). {The ergodic fading channel, in which each transmitted codeword experiences the entire statistics of the fading process, is commonly assumed to analyze the theoretical performance of FSO systems \cite{nistazakis2009average,gappmair2011further}.} Moreover, we assume the fading gains are independent RVs. This can be easily achieved by placing the apertures a few centimeters apart, due to the fact that the coherence length of the irradiance fluctuations is of the order of centimeters \cite{navidpour2007ber,lee2004part}.

\subsection{Electrical SNR}\label{se:E_SNR}
We normalized the average gains $\bar{I}_{mn} \triangleq {\text E}{\left(I_{mn}\right)}$ at each branch to unit. It has been shown in \cite{bayaki2009performance} that in turbulence fading channels, the performance gains provided by the complicated MRC scheme are only small to moderate when compared with the simple EGC scheme. Therefore, EGC is more attractive in FSO systems due to its remarkable lower implementation complexity. When the EGC method is used, the received sum output signal is
\begin{align}\label{eq:EGC}
 {y   = \frac{\sqrt{P}{\eta s}}{{MN}}I + v,}
\end{align}
where $I \triangleq \sum\limits_{n = 1}^N {\sum\limits_{m = 1}^M {{I_{mn}}} } $, {and $v$ is the effective Gaussian noise at the receiver. We normalized the total transmit power $P=M{P_\text{avg}}=1$.} Note that the division by $MN$ in \eqref{eq:EGC} is used to ensure that the total transmit power and the sum of the receiver aperture areas are the same as that of a system with single transmit/receive antenna, respectively. Then, the instantaneous electrical SNR of the sum output signal at the receiver can be calculated as
\begin{align}\label{eq:EGC_SNR}
{\gamma  } = \frac{{{{\left( {\eta I} \right)}^2}}}{{{{\left( {MN} \right)}^2}{N_0}}}=\gamma_0 I^2,
\end{align}
{where $\gamma_0 \triangleq \frac{{{{ {\eta }  }^2}}\rho}{{{{\left( {MN} \right)}^2} }}$ with $\rho=1/{N_0}$ denoting the transmit SNR}, and the average electrical SNR can be correspondingly denoted by ${\bar{\gamma} } = \gamma_0 {\bar{I}}^2$.

\subsection{$\Gamma \Gamma$ channel model}\label{se:gamma_model}
Adopting suitable model for the turbulence fading channel is of key importance to assess the performance of FSO systems. It has been proved in \cite{andrews2005laser,ghassemlooy2012optical} that the $\Gamma \Gamma$ distribution can well characterize the fading gains $I_{mn}$ for turbulence scenarios from weak to strong. Based on a doubly stochastic theory of scintillation, the small-scale irradiance fluctuations are modulated by large-scale irradiance fluctuations of the propagating wave. Then, the $\Gamma \Gamma$ distribution can be derived from the product of two independent Gamma distributions and the PDF of $I_{mn}$ is given by \cite{chatzidiamantis2011distribution}
\begin{align}\label{eq:gamma_gamma}
{f_{{I_{mn} }}}\left( {{I_{mn}}} \right) &= \frac{{2{{\left( {a b } \right)}^{\frac{{a  + b }}{2}}}}}{{\Gamma \left( a  \right)\Gamma \left( b  \right){{\bar I}_{mn}}}}{\left( {\frac{{{I_{mn}}}}{{{{\bar I}_{mn}}}}} \right)^{\frac{{a  + b }}{2} - 1}}\notag \\
&\qquad \times {{\rm K}_{a  - b }}\left( {2\sqrt {\frac{{a b }}{{{{\bar I}_{mn}}}}{I_{mn}}} } \right),
\end{align}
where $a$ and $b$ are the shaping parameters of small-scale and large-scale eddies of the scattering environment, $\Gamma\left(\cdot \right)$ denotes the Gamma function \cite[Eq. 8.310.1]{gradshtein2000table}, and ${\rm K}_v \left(\cdot \right)$ is related to the modified Bessel function of the second kind with order $v$ \cite[Eq. 9.6.1]{abramowitz1964handbook}. {In this paper, we assume the spherical wave propagation, which is the exact model for line-of-sight MIMO systems \cite{Bohagen2010on} and widely adopted in the literature \cite{nistazakis2009average,uysal2004error,uysal2006error,deng2013capacity,antonio2010capacity}.} The effective numbers $a$ and $b$ are related to the atmospheric conditions and are respectively given by \cite{bayaki2009performance}
\begin{align}
a  &= {\left( {\exp \left[ {\frac{{0.49\sigma _2^2}}{{{{\left( {1 + 0.18{d^2} + 0.56\sigma _2^{12/5}} \right)}^{7/6}}}}} \right] - 1} \right)^{ - 1}},\label{eq:beta_OC} \\
b &= {\left( {\exp \left[ {\frac{{0.51\sigma _2^2{{\left( {1 + 0.69\sigma _2^{12/5}} \right)}^{ - 5/6}}}}{{{{\left( {1 + 0.9{d^2} + 0.62{d^2}\sigma _2^{12/5}} \right)}^{5/6}}}}} \right] - 1} \right)^{ - 1}},\label{eq:m_OC}
\end{align}
where $\sigma _2^2 = 0.492C_n^2{\hat{k}^{7/6}}{L^{11/6}}$ is the Rytov variance and $d = \sqrt {\hat{k}{D^2}/4L} $ with $L$ being the distance between transmitter and receiver, $\hat{k} \triangleq 2\pi/\lambda$ denotes the optical wavenumber with $\lambda$ being the operational wavelength, $D$ is the aperture diameter of the receiver, and $C_n^2$ is the altitude-dependent index of the refractive
structure parameter determining the turbulence strength. Furthermore, we assume $C_n^2$ remains constant for relatively long transmit bits interval and varies from $10^{-17} {\text{m}}^{-2/3}$ to $10^{-13} {\text{m}}^{-2/3}$ for weak to strong turbulence cases \cite{navidpour2007ber}.

According to the experimental data, we can properly choose the values of $a$ and $b$ in \eqref{eq:gamma_gamma} to provide a good approximation of the irradiance fluctuation PDF. Moreover, $a$ and $b$ are linked to the scintillation index, which is used to describe the strength of atmospheric fading and given by \cite[Eq. (4)]{antonio2010capacity}
\begin{align}\label{eq:SI}
\text{S.I.} \triangleq \frac{\text{E}{(I^2)}}{[\text{E}{(I)}]^2}-1= \frac{1}{a} + \frac{1}{b} + \frac{1}{{ab}}.
\end{align}
From \eqref{eq:SI}, it is clear to see that the scintillation index depends on the values of $a$ and $b$. Furthermore, the turbulence becomes weaker for decreasing scintillation index as $a$ and/or $b$ increase, and gets stronger for increasing scintillation index with smaller values of $a$ and/or $b$.
The $q$th moment of $I_{mn}$ can be obtained by using \cite[Eq. (6.561.16)]{gradshtein2000table} as
\begin{align}\label{eq:moment_I}
{\text E}\left( {{I_{mn}^q}} \right) = \frac{{\Gamma \left( {a  + q} \right)\Gamma \left( {b  + q} \right)}}{{\Gamma \left( a  \right)\Gamma \left( b  \right)}}{\left( {\frac{{a b }}{{\bar I}}} \right)^{ - q}}.
\end{align}

\section{Ergodic Capacity Analysis}\label{se:MIMO}
{Considering the number of transmit apertures is less than the number of receive apertures, e.g., $M<N$, the ergodic capacity of MIMO FSO system in bits/s/Hz is given by \cite{matthaiou2010capacity}}
\begin{align}\label{eq:capacity}
{\text C} =  {\text{E}}\Bigg( {{{\log }_2}\left[ {\det \left( {{{\pmb{I}}_M} + \frac{\gamma }{M}{\pmb{H}}^H{{\pmb{H}} }} \right)} \right]} \Bigg),
\end{align}
where $\pmb{I}_M$ denotes an $M\times M$ identity matrix, and $\pmb{H}$ is an $N\times M$ channel matrix with elements being $I_{nm}$. In this section, we present some statistical features of the sum of $L\triangleq MN$ i.i.d. and i.n.i.d. $\Gamma \Gamma$ RVs $I_l$, where $S\triangleq \sum\limits_{l = 1}^L {{I_l}}$. The approximation methods for the PDF of the sum of $\Gamma \Gamma$ RVs are introduced\footnote{The PDF analysis can also be studied by the method of generalized power series representation \cite{park2011average}, which induces an infinite series PDF expression.}. These results are useful for the ergodic capacity analysis of MIMO FSO systems over turbulence channels.

\subsection{I.i.d. $\Gamma \Gamma$ Fading Channels}
{Recently, the $\alpha$-$\mu$ distribution has been proposed to approximate the sum of multiple i.i.d. $\Gamma \Gamma$ RVs \cite{peppas2011simple,yang2012appro}.} This is because the $\alpha$-$\mu$ distribution is a more general distribution, which includes the Gamma distribution as a special case. We use $R$ to denote an $\alpha$-$\mu$ RV. Then, the PDF expression of $\alpha$-$\mu$ distribution is given by \cite{yacoub2007alpha}
\begin{align}\label{eq:PDF_R}
{f_R}\left( r \right) &= \frac{{\alpha {\mu ^\mu }{r^{\alpha \mu  - 1}}}}{{{{\hat r}^{\alpha \mu }}\Gamma \left( \mu  \right)}}\exp \left( { - \frac{{\mu {r^\alpha }}}{{{{\hat r}^\alpha }}}} \right),
\end{align}
where $\alpha> 0$, $\mu = {{\text E}^2}\left( {{R^\alpha }} \right)/{\text {Var}}\left( {{R^\alpha }} \right)$ is the inverse of the normalized variance of $R^\alpha$, ${\hat r}=\left[{\text E}\left(R^\alpha\right)\right]^{1/\alpha}$ is a $\alpha$-root mean.

With the knowledge of the first, the second and the fourth moment of $R$, the moment-matching method can be used to approximate the parameters $\alpha$ and $\mu$ \cite{yacoub2007alpha}. Therefore, we need to derive the parameters $\alpha$, $\mu$ and $\hat r$ by solving the following nonlinear functions:
\begin{align}\label{eq:moment_match}
\frac{{{{\text E}^2}\left( I \right)}}{{{\text E}\left( {{I^2}} \right) - {{\text E}^2}\left( I \right)}} &= \frac{{{\Gamma ^2}\left( {\mu  + 1/\alpha } \right)}}{{\Gamma \left( \mu  \right)\Gamma \left( {\mu  + 2/\alpha } \right) - {\Gamma ^2}\left( {\mu  + 1/\alpha } \right)}},\notag \\
\frac{{{{\text E}^2}\left( {{I^2}} \right)}}{{{\text E}\left( {{I^4}} \right) - {{\text E}^2}\left( {{I^2}} \right)}} &= \frac{{{\Gamma ^2}\left( {\mu  + 2/\alpha } \right)}}{{\Gamma \left( \mu  \right)\Gamma \left( {\mu  + 4/\alpha } \right) - {\Gamma ^2}\left( {\mu  + 2/\alpha } \right)}},\notag \\
\hat r &= \frac{{{\mu ^{1/\alpha }}\Gamma \left( \mu  \right){\text E}\left( I \right)}}{{\Gamma \left( {\mu  + 1/\alpha } \right)}},
\end{align}
where the moments in \eqref{eq:moment_match} can be evaluated by using \eqref{eq:moment_I} and the multinomial identity as \cite[Eq. (10)]{peppas2011simple}
\begin{align}\label{eq:multinomial_identity}
{\text E}\left( {{I^q}} \right) &= \sum\limits_{{j_1} = 0}^q {\cdots\sum\limits_{{j_{l - 1}} = 0}^{{j_{l - 2}}} {\left( {\begin{array}{*{20}{c}}
q\\
{{j_1}}
\end{array}} \right)} }\cdots\left( {\begin{array}{*{20}{c}}
{{j_{l - 2}}}\\
{{j_{l - 1}}}
\end{array}} \right)\notag \\
&\qquad \times {\text E}\left( {I_{1}^{q - {j_1}}} \right)\cdots{\text E}\left( {I_{L}^{{j_{l - 1}}}} \right).
\end{align}
Note that an analytical solution to the equations \eqref{eq:moment_match} is very difficult to be obtained, so we use numerical methods instead, such as the \emph{fsolve} function of Matlab and Maple.

{Based on \eqref{eq:capacity}, the ergodic capacity of MIMO FSO systems with EGC can be expressed as}
\begin{align}\label{eq:capacity_egc}
{\text C}  = \int_0^\infty  {{{\log }_2}\left( {1 + \gamma} \right)f\left( \gamma \right)d\gamma}.
\end{align}
Using a simple power transformation of \eqref{eq:EGC_SNR}, we have
\begin{align}\label{eq:power_trans}
{f\left( \gamma  \right) = f_R{\left( \gamma \right)}\frac{{dI}}{{d\gamma }}.}
\end{align}
Recall \eqref{eq:PDF_R}, the PDF of the instantaneous electrical SNR $\gamma$ at the receiver side can be given by
\begin{align}\label{eq:pdf_egc_snr}
{f_{\text {i.i.d.}}}\left( \gamma  \right) = \frac{{\alpha {\mu ^\mu }{\gamma ^{\frac{{\alpha \mu }}{2} - 1}}}}{{2\gamma _0^{\frac{{\alpha \mu }}{2}}{{\hat I}^{\alpha \mu }}\Gamma \left( \mu  \right)}}\exp \left( { - \frac{{\mu {\gamma ^{\frac{\alpha }{2}}}}}{{{{\hat I}^\alpha }\gamma _0^{\frac{\alpha }{2}}}}} \right).
\end{align}
{where ${\hat I}$ is the $\alpha$-root mean value of the approximation.} Then, \eqref{eq:capacity_egc} can be rewritten as
\begin{align}\label{eq:capacity_egc1}
{\text C}_{\text {i.i.d.}}  &= \frac{{\alpha {\mu ^\mu }}}{{2\gamma _0^{\alpha \mu /2}{{\hat I}^{\alpha \mu }}\Gamma \left( \mu  \right)\ln 2}}\notag \\
&\times \int_0^\infty  {{\gamma ^{\alpha \mu /2 - 1}}\ln \left( {1 + \gamma } \right)\exp \left( { - \frac{{\mu {\gamma ^{\alpha /2}}}}{{{{\hat I}^\alpha }\gamma _0^{\alpha /2}}}} \right)d\gamma}.
\end{align}

To evaluate \eqref{eq:capacity_egc1}, we can express the logarithmic and exponential functions as Meijer's $G$-functions by \cite[Eq. (11)]{adamchik1990algorithm}
\begin{align}\label{eq:log_meijer}
\ln \left( {1 + x} \right) = G_{2,2}^{1,2}\left[{ {x} \left|  \begin{array}{{c}}
{1,1}\\
{1,0}\\
\end{array} \right.} \right], {e^{ - x}} = G_{0,1}^{1,0}\left[ {x\left| {\begin{array}{*{20}{c}}
 - \\
0
\end{array}} \right.} \right].
\end{align}
Upon using \cite[Eq. (21)]{adamchik1990algorithm}, the analytical expression of ergodic capacity is given by
\begin{align}\label{eq:capacity_egc_result}
&{\text C}_{\text {i.i.d.}}= \frac{{\alpha {\mu ^\mu }\sqrt k }}{{2l\gamma _0^{\frac{{\alpha \mu }}{2}}{{\hat I}^{\alpha \mu }}\Gamma \left( \mu  \right){{\left( {2\pi } \right)}^{l + k/2 - 3/2}}\ln 2}}\notag \\
&\times G_{2l,k + 2l}^{k + 2l,l}\left[ {{{\left( {\frac{\mu }{{k{{\hat I}^\alpha }\gamma _0^{\frac{\alpha }{2}}}}} \right)}^k}\left| {\begin{array}{*{20}{c}}
{\Delta \left( {l, - \frac{{\alpha \mu }}{2}} \right),\Delta \left( {l,1 - \frac{{\alpha \mu }}{2}} \right)}\\
{\Delta \left( {k,0} \right),\Delta \left( {l, \frac{{-\alpha \mu }}{2}} \right),\Delta \left( {l, \frac{{- \alpha \mu }}{2}} \right)}
\end{array}} \right.}\! \right]
\end{align}
where $\Delta \left( {\epsilon, \tau } \right)=\frac{\tau }{\epsilon}, \frac{\tau+1 }{\epsilon},\cdots, \frac{\tau +\epsilon -1}{\epsilon}$, with $\tau$ is an arbitrary real value and $\epsilon$ is a positive integer. Moreover, $l/k=\alpha/2$, where $l$ and $k$ are both positive integers. For example, if $\alpha= 0.8$, we can set $l=2$ and $k=5$. Furthermore, $l=\alpha$ and $k=2$ are for the special case of $\alpha \in {\mathbb{Z}}^+$. Note that the calculation of \eqref{eq:capacity_egc_result} is efficient for special values of $\alpha$. However, for large values of $l$ and $k$, it is very tedious to use \eqref{eq:capacity_egc_result}. Therefore, another method is adopted to solve this problem as follows.

We recall the well-known translation from the Meijer's $G$-function to the Fox's $H$-function as \cite[Eq. (8.3.2.21)]{prudnikov1990integrals}
\begin{align}
H_{p,q}^{m,n}\left[ {x\left| {\begin{array}{*{20}{c}}
{\left[ {{a_p},1} \right]}\\
{\left[ {{b_p},1} \right]}
\end{array}} \right.} \right] = G_{p,q}^{m,n}\left[ {x\left| {\begin{array}{*{20}{c}}
{\left[ {{a_p}} \right]}\\
{\left[ {{b_p}} \right]}
\end{array}} \right.} \right].
\end{align}
With the help of \eqref{eq:log_meijer}, the logarithmic and exponential functions can be expressed in terms of the Fox's $H$-function as
\begin{align}
\ln \left( {1 + x} \right) &= H_{2,2}^{1,2}\left[ {x\left| {\begin{array}{*{20}{c}}
{\left( {1,1} \right),\left( {1,1} \right)}\\
{\left( {1,1} \right),\left( {0,1} \right)}
\end{array}} \right.} \right],\label{eq:log_H}\\
{e^ { - x} } &= H_{0,1}^{1,0}\left[ {x\left| {\begin{array}{*{20}{c}}
 - \\
{\left( {0,1} \right)}
\end{array}} \right.} \right]. \label{eq:exp_H}
\end{align}
Substituting \eqref{eq:exp_H} and \eqref{eq:log_H} into \eqref{eq:capacity_egc1}, we get the Mellin--Barnes integral \cite{Paris2001Asymptotics} of the product of two Fox's $H$-functions. By involving \cite[Eq. (2.25.1.1)]{prudnikov1990integrals}, the ergodic capacity of MIMO FSO system with EGC scheme for arbitrary values of $\alpha$ and $\mu$ is given by
\begin{align}\label{eq:capacity_egc_result_H}
{\text C}_{\text {i.i.d.}}  = \frac{1}{{\Gamma \left( \mu  \right)\ln 2}}H_{3,2}^{1,3}\left[ {\frac{{{{\hat I}^2}{\gamma _0}}}{{{\mu ^{2/\alpha }}}}\left| {\begin{array}{*{20}{c}}
{\left( {1,1} \right),\left( {1,1} \right),\left( {1 - \mu ,2/\alpha } \right)}\\
{\left( {1,1} \right),\left( {0,1} \right)}
\end{array}} \right.} \right].
\end{align}

{Note that \eqref{eq:capacity_egc_result_H} is very compact in the form of Fox's $H$-functions and will be verified through Monte-Carlo simulations in Section \ref{se:numerical_results}. We can use the numerically efficient methods presented in \cite{peppas2012ol,yilmaz2009product} to evaluate the Fox's $H$-function.} The path of the integration is running from $\omega-i\infty$ to $\omega+i\infty$, which depend on all poles of gamma functions in the numerator \cite{carter1977distribution}.
{The ergodic capacity expression of SISO FSO links has been given in \cite[Eq. (16)]{nistazakis2009average} by using the PDF of $\Gamma\Gamma$ distribution directly. However, the $\alpha$-$\mu$ approximation method is utilized in our work to study the MIMO FSO system. It should be pointed out that although \eqref{eq:capacity_egc_result_H} cannot intuitively reduce to the existing expression for the SISO case (e.g., \cite[Eq. (16)]{nistazakis2009average}), their mathematical calculating results are quite close to each other, which will be verified by the simulation results in Fig. 1 and 2 ($M=N=1$ for the SISO link).}

\newcounter{mytempeqncnt}
\begin{figure*}[!b]
\normalsize
\setcounter{mytempeqncnt}{\value{equation}}
\setcounter{equation}{29}
\hrulefill
\begin{align}\label{eq:inid_capacity_expression}
{\text C}_{\text {i.n.i.d.}} =&
  \sum_{i=1}^L \sum_{j=1}^{m_i}\frac{1}{4\pi\Gamma(Lk)\Gamma(j)\ln2} \left(\frac{Lkm_i}{\Omega_i\gamma_0^{\frac{1}{2}}}\right)^{\frac{Lk+j}{2}}w_L\left(i,j,\{m_l\}_{l=1}^L,\{\Omega_l\}_{l=1}^L\right)\notag\\
  &\times G_{2,6}^{6,1}\left(\frac{1}{16\gamma_0}\left(\frac{Lkm_i}{\Omega_i}\right)^2\left|
    \begin{array}{c}
    -\frac{Lk+j}{4},1-\frac{Lk+j}{4}\\
    \frac{Lk-j}{4},\frac{Lk-j+2}{4},-\frac{Lk-j}{4},-\frac{Lk-j-2}{4},-\frac{Lk+j}{4},-\frac{Lk+j}{4}
    \end{array}\right.
    \right).
\end{align}
\setcounter{equation}{\value{mytempeqncnt}}
\vspace*{2pt}
\end{figure*}
\vspace{-0.1cm}
\subsection{I.n.i.d. $\Gamma \Gamma$ Fading Channels}
The ergodic capacity for i.i.d. $\Gamma \Gamma$ fading channels has been derived in \eqref{eq:capacity_egc_result_H}, now we move on to the ergodic capacity analysis for the i.n.i.d. $\Gamma \Gamma$ fading channels. {It is reasonable to assume i.n.i.d. fading when the transmit or receive apertures are placed far from the others.} The approximative PDF expression for the sum of multiple i.n.i.d. $\Gamma\Gamma$ RVs has been presented in \cite{chatzidiamantis2011distribution} by neglecting the cross terms as follows
\begin{align}\label{eq:inid_approx}
  S=\sum\limits_{l = 1}^L {{I_l}}=\sum_{l=1}^L x_ly_l \approx \frac{1}{L}\left(\sum_{l=1}^L x_l\right)\left(\sum_{l=1}^L y_l\right),
\end{align}
where $x_l$ and $y_l$ are Gamma RVs with parameters $(k_l,1/k_l)$ and $(m_l,\Omega_l/m_l)$ respectively, and $I_l=x_ly_l$ is a decomposition from a $\Gamma\Gamma$ RV to two independent Gamma RVs. {Following the well-known property for the product of two random variables \cite{matthaiou2010capacity}, the PDF of $I_l$ can be given in the form of \eqref{eq:gamma_gamma} with the shape parameters $k_l=a$, $m_l=b$, and the mean power $\Omega_l= {{{\bar I}_{mn}}}$.} {In most practical MIMO FSO systems, the transmit lasers and receive apertures are placed only a few centimeters apart. Then the large-scale irradiance fluctuation changes relatively slowly compared with the small-scale effect. It is reasonable to assume that the $\Gamma\Gamma$ RVs have one shaping parameter in a column, i.e., $k_l=k$.} Based on (\ref{eq:inid_approx}), the PDF for the sum of $L$ i.n.i.d. $\Gamma\Gamma$ RVs is shown in \cite[Eq. (27)]{chatzidiamantis2011distribution} as
\begin{align}\label{eq:inid_sum_pdf}
  f_S&(s) = \sum_{i=1}^L\sum_{j=1}^{m_i}w_L\left(i,j,\{m_l\}_{l=1}^L,\{\Omega_l\}_{l=1}^L\right)\notag\\
  &\times \frac{2(Lkm_i)^{\frac{Lk+j}{2}}s^{\frac{Lk+j}{2}-1}}{\Gamma(Lk)\Gamma(j)\Omega_i^{\frac{Lk+j}{2}}}
  \text{K}_{Lk-j}\left(2\left(\frac{Lkm_i}{\Omega_i}s\right)^{\frac{1}{2}}\right),\notag
\end{align}
where the weights can be easily computed by the recursive formula according to {\cite[Eq. (24)]{chatzidiamantis2011distribution}}
\begin{align}
  w_L\left(i,\right.&m_i-t,\left.\{m_l\}_{l=1}^L,\{\Omega_l\}_{l=1}^L\right)\notag\\
  &=\frac{1}{t}\sum_{q=1,q\neq i}^L\sum_{j=1}^t m_q\left(1-\frac{\Omega_im_q}{\Omega_qm_i}\right)^{-j}\notag\\
  &\times w_L\left(i,m_i-t+j,\{m_l\}_{l=1}^L,\{\Omega_l\}_{l=1}^L\right),
\end{align}
where $t=1,\cdots,m_i-1$, and
\begin{align}
  w_L\left(i,m_i,\{m_l\}_{l=1}^L,\{\Omega_l\}_{l=1}^L\right)=
  \prod_{j=1,j\neq i}^L\left(1-\frac{\Omega_jm_i}{\Omega_im_j}\right)^{-m_j}.\notag
\end{align}

{For MIMO FSO systems with EGC receiver, we can approximate the PDF of the instantaneous electrical SNR $\gamma$ when one of the shaping parameters remains the same with a nested finite weighted sum of $\Gamma\Gamma$ PDFs as follow:
\begin{align}\label{eq:inid_sum_weighted_pdf}
&{f_\text {i.n.i.d.}}\left( \gamma  \right)= \sum\limits_{i = 1}^L {\sum\limits_{j = 1}^{{m_i}} {{\omega _L}\left( {i,j,\{ {m_l}\} _{l = 1}^L,\{ {\Omega _l}\} _{l = 1}^L} \right)} } \notag \\
&\times \frac{{{{\left( {Lk{m_i}} \right)}^{\frac{{Lk + j}}{2}}}{{\sqrt \gamma  }^{\frac{{Lk + j}}{2}}}}}{{\Gamma \left( j \right)\Gamma \left( {Lk} \right){{\left( {{\Omega _i}\sqrt {{\gamma _0}} } \right)}^{\frac{{Lk + j}}{2}}}}}{K_{Lk - j}}\left( {2\sqrt {\frac{{Lk{m_i}\gamma }}{{{\Omega _i}{\gamma _0}}}} } \right).
\end{align}}
{Note that the accuracy of \eqref{eq:inid_sum_weighted_pdf} depends on the combinations of the parameters $k$, $m_l$, and $\Omega_l$. This constraint has been explicitly investigated by employing Kolmogorov-Smirnov goodness-of-fit test in Section III. C of \cite{chatzidiamantis2011distribution}.}


For i.n.i.d. $\Gamma \Gamma$ turbulence channels, the ergodic capacity in (\ref{eq:capacity_egc}) can be written as
\begin{align}\label{eq:inid_capacity}
  &{\text C}_{\text {i.n.i.d.}} = \sum_{i=1}^L \sum_{j=1}^{m_i}c_{ij}w_L\left(i,j,\{m_l\}_{l=1}^L,\{\Omega_l\}_{l=1}^L\right)\notag\\
  &\times\int_0^\infty \ln(1\!+\!\gamma)\gamma^{\frac{Lk\!+\!j}{4}\!-\!1}\text{K}_{Lk\!-\!j}\left(2\left(\frac{Lkm_i\sqrt{\gamma}}
  {\Omega_i\sqrt{\gamma_0}}\right)^{\frac{1}{2}}\right)\text{d}\gamma,
\end{align}
where $c_{ij}$ is a constant dependent on $i,j$, and given by
\begin{align}\label{eq:inid_capacity_const}
  c_{ij}\triangleq\frac{1}{\Gamma(Lk)\Gamma(j)\ln2} \left(\frac{Lkm_i}{\Omega_i\sqrt{\gamma_0}}\right)^{\frac{Lk+j}{2}}.
\end{align}
Using \cite[Eq. (21)]{adamchik1990algorithm}, the integral in (\ref{eq:inid_capacity}) can be expressed as
\begin{align}\label{eq:inid_capacity_int}
 &\int_0^\infty \ln(1+\gamma)\gamma^{u-1}\text{K}_{v}\left(2\sqrt{z\gamma^{1/2}}\right)\text{d}\gamma\notag\\
 &= \frac{1}{2 }\int_0^\infty \gamma^{u-1}G_{2,2}^{1,2}\left[\gamma\left|\begin{array}{c}
      1,1 \\
      1,0
    \end{array}\right.
    \right]
     G_{0,2}^{2,0}\left[z\sqrt{\gamma}\left|\begin{array}{c}
      - \\
      \frac{v}{2},\frac{-v}{2}
    \end{array}\right.
    \right]\text{d}\gamma\notag\\
 &= \frac{1}{4\pi} G_{2,6}^{6,1}\left[\frac{z^2}{16}\left|
    \begin{array}{c}
    -u,1-u\\
    \frac{v}{4},\frac{v+2}{4},-\frac{v}{4},-\frac{v-2}{4},-u,-u
    \end{array}\right.
    \right],
\end{align}
where we have used the notations of $u \triangleq {(Lk+j)}/{4}$, $ v \triangleq Lk-j$, $z \triangleq {Lkm_i}/({\Omega_i\sqrt{\gamma_0}})$, and the identity \cite[Eq. 9.34.3]{gradshtein2000table}
\begin{align}
  \text{K}_v(x) = \frac{1}{2}G_{0,2}^{2,0}\left[\frac{x^2}{4}\left|\begin{array}{c}
      - \\
      \frac{v}{2},-\frac{v}{2}
    \end{array}\right.
    \right].
\end{align}
By substituting \eqref{eq:inid_capacity_int} into \eqref{eq:inid_capacity}, the ergodic capacity of MIMO FSO systems over i.n.i.d. $\Gamma\Gamma$ fading channels is expressed as \eqref{eq:inid_capacity_expression} at the bottom of this page.


\subsection{High-SNR Analysis}
In order to get more insights into the impact of system and channel parameters on the ergodic capacity, we investigate the asymptotic ergodic capacity of FSO MIMO systems in the high-SNR regime. First, we consider the i.i.d. case. By taking $\gamma$ large in \eqref{eq:capacity_egc1}, the ergodic capacity at high SNRs becomes
\setcounter{equation}{30}
\begin{align}\label{eq:capacity_up_high}
{\text C}_{\text {i.i.d.}} ^\infty &=  \frac{1}{{\ln 2}} {\int_0^\infty  {\ln \left( { \gamma} \right)} \frac{{\alpha {\mu ^\mu }{\gamma ^{\frac{{\alpha \mu }}{2} - 1}}}}{{2\gamma _0^{\frac{{\alpha \mu }}{2}}{{\hat I}^{\alpha \mu }}\Gamma \left( \mu  \right)}}\exp \left( { - \frac{{\mu {\gamma ^{\frac{\alpha }{2}}}}}{{{{\hat I}^\alpha }\gamma _0^{\frac{\alpha }{2}}}}} \right)d\gamma}.
\end{align}
With the help of \cite[Eq. (2.6.21.2)]{prudnikov1986integrals}
\begin{align}\label{eq:intergral}
&\int_0^\infty  {{x^{\zeta  - 1}}\exp \left( { - p{x^\theta }} \right)} \ln{x} dx  \notag \\
&\qquad= {\theta ^{ - 2}}{p^{ -\zeta /\theta }}\Gamma \left( {\zeta /\theta } \right)\left[ {\psi \left( {\zeta /\theta } \right) - \ln p} \right],
\end{align}
\eqref{eq:capacity_up_high} can be further calculated as
\begin{align}\label{eq:capacity_up_high_result}
{\text C}_{\text {i.i.d.}} ^\infty &= \frac{2}{{\alpha \ln 2}}\left[ {\psi \left( \mu  \right) - \ln \mu  + \alpha \ln \hat I + \frac{\alpha }{2}\ln {\gamma _0}} \right],
\end{align}
where $\psi\left( \cdot  \right)$ is the Euler's digamma function \cite[Eq. (8.360.1)]{gradshtein2000table}.
It is obvious that the high-SNR ergodic capacity increases with the optical-electrical conversion coefficient $\eta$. Note that similar observations were also made in \cite{antonio2010capacity,luong2013average}.

{Now we consider the high-SNR ergodic capacity of MIMO FSO systems over i.n.i.d. $\Gamma\Gamma$ fading channels. Using the similar method aforementioned, we approximate $\ln(1+\gamma)$ with $\ln(\gamma)$ by taking $\gamma$ large in the i.n.i.d. case, so the integral in (\ref{eq:inid_capacity_int}) can be expressed as
\begin{align}\label{eq:inid_capacity_int_app}
&\int_0^\infty \ln(\gamma)\gamma^{u-1}\text{K}_{v}\left(2\sqrt{z\gamma^{1/2}}\right)\text{d}\gamma=\frac{2}{{{z^{2u}}}}\Gamma \left( {\frac{{4u + v}}{2}} \right)\notag\\
 &\times \Gamma \left( {\frac{{4u - v}}{2}} \right)\left[\psi \left( {\frac{{4u + v}}{2}} \right) + \psi \left( {\frac{{4u - v}}{2}} \right) - \ln \left( z \right)\right],
\end{align}
where we have used the identity in \cite[Eq. (2.16.20.1)]{prudnikov1992integrals}. Substituting \eqref{eq:inid_capacity_int_app} into \eqref{eq:inid_capacity} and after some basic algebraic manipulations, the {analytical expression} for the high-SNR capacity can be finally derived as
\begin{align}\label{eq:inid_capacity_expression_app}
{\text C}_{\text{i.n.i.d.}}^\infty  &= \frac{2}{{\ln 2}}\sum\limits_{i = 1}^L \sum\limits_{j = 1}^{{m_i}} {\omega _L}\left( {i,j,\{ {m_l}\} _{l = 1}^L,\{ {\Omega _l}\} _{l = 1}^L} \right)\notag \\
& \times \left[ {\psi \left( {Lk} \right) + \psi \left( j \right) - \ln \left( {Lk{m_i}} \right) + \ln \left( {{\Omega _i}\sqrt {{\gamma _0}} } \right)} \right].
\end{align}
The above expression is intuitive, since it indicates that at high SNRs, the larger fading power $\Omega_i$, transmit SNR $\rho$, and the optical-electrical conversion coefficient $\eta$ can achieve higher ergodic capacity of MIMO FSO systems over i.n.i.d. $\Gamma\Gamma$ turbulence channels.}


\section{Numerical Results}\label{se:numerical_results}

In this section, the analytical results of MIMO FSO systems over different system configuration scenarios and/or various channel fading conditions are presented and compared with Monte-Carlo simulations. The impact of some system and channel parameters on the ergodic capacity is also analyzed in detail. {Moreover, the ergodic capacity for SISO FSO systems is also provided as a benchmark for comparison.} We assume a wavelength of $\lambda=850$ nm and an optical-to-electrical conversion coefficient $\eta=1$ without loss of generality. In Monte-Carlo simulations, $10^7$ i.i.d. and/or i.n.i.d. $\Gamma \Gamma$ random samples are generated. In all the cases considered here, there is a good match between the analytical and the respective simulated results, which validates the accuracy of the proposed expressions.

\begin{figure}[tbp]
\centering
\includegraphics[scale=0.65]{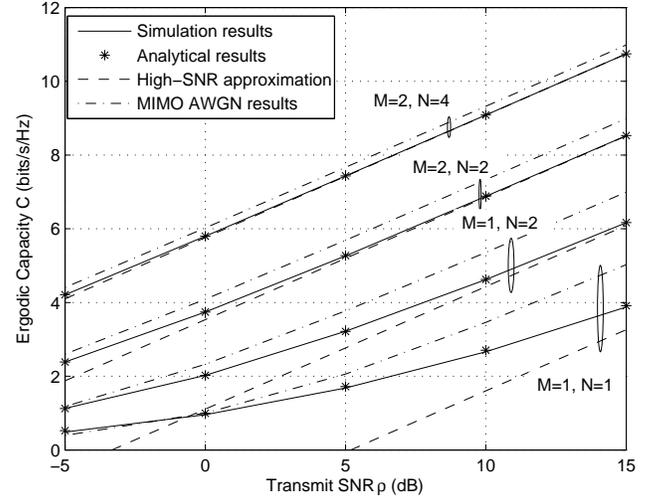}
\caption{Simulated, analytical, and high-SNR approximation ergodic capacity of MIMO FSO systems over i.i.d. distributed strong turbulence fading channels versus the transmit SNR ${{\rho }} $ ($C_n^2= 3\times 10^{-14} {\text{m}}^{-2/3}$, $D=0.01$ m, $L=4000$ m, $\lambda=850$ nm and $\eta = 1$). Moreover, the capacity of the non-turbulent channel, e.g., MIMO AWGN channel, is plotted to provide a benchmark for comparison.
\label{fig:Capacity_M_N}}
\end{figure}

{Figure \ref{fig:Capacity_M_N} portrays the simulated, analytical \eqref{eq:capacity_egc_result_H} and high-SNR \eqref{eq:capacity_up_high_result} approximated ergodic capacity of MIMO FSO systems as a function of the average electrical SNR ${{\bar \gamma  }} $ for various values of $M$ and $N$.} We assume the strong turbulence fading channels of the MIMO FSO system are i.i.d., with fix values of $C_n^2= 3\times 10^{-14} {\text{m}}^{-2/3}$, $D=0.01$ m, and $L=4000$ m. For comparison, we also plot the ergodic capacity of MIMO FSO systems over AWGN channel, whose ideal capacity is $\text{C}_\text{AWGN} = \log_2{(1 + \text{SNR})}$, as the benchmark for comparison. It is clear to see that the analytical ergodic capacity is very accurate in the entire SNR regime, and the high-SNR approximation curves are quite tight in the moderate and high SNRs. As expected, the ergodic capacity is significantly improved as the number of transmit and receive apertures increases, especially in high-SNR regime. Finally, the gap between the MIMO AWGN and analytical curves decreases as $M$ and $N$ get larger. When $M=2$ and $N=4$, the analytical ergodic capacity is very close to the capacity of the MIMO AWGN channel.

Recall that the MIMO FSO systems favor high data rate with broad spectrum, it is significant to achieve an accurate ergodic capacity approximation to guide the practical system design. In order to emphasize the accuracy of our proposed results, the absolute difference between the analytical and simulated ergodic capacity is presented in Table \ref{tab:addlabel}. Moreover, the approximation errors of the ergodic capacity deduced in \cite{deng2013capacity} are also presented for comparison. From Table \ref{tab:addlabel}, it is clear to see that our approximation outperforms the results proposed in \cite{deng2013capacity} in terms of approximation error. Considering the typical case of $M=N=2$, $\bar{\gamma}=1$ dB, our analytical result is near hundredfold accurate compared with \cite{deng2013capacity}. Moreover, the approximation error tends to ascend as the average electrical SNR increases when using the method proposed in \cite{deng2013capacity}, while the approximation error of our $\alpha$-$\mu$ approximation is stable and small in the entire SNR regime.

\begin{table*}[htbp]
  \centering
  \caption{Performance Error Comparison between Our Results and the Results Presented in \cite{deng2013capacity} in bit/s/Hz ($C_n^2= 3\times 10^{-14} {\text{m}}^{-2/3}$, $D=0.01$ m, $L=4000$ m, $\lambda=850$ nm and $\eta = 1$)}
  \begin{threeparttable}
    \begin{tabular}{|c|c|c|c|c|c|c|}
    \hline
    \multirow{2}[4]{*}{$\bar{\gamma}$(dB)} & \multicolumn{3}{c|}{Our Results} & \multicolumn{3}{c|}{Results in \cite{deng2013capacity}} \bigstrut\\
\cline{2-7}          & $M=1, N=2$ & $M=2, N=2$ & $M=2, N=4$ & $M=1, N=2$ & $M=2, N=2$ & $M=2, N=4$ \bigstrut\\
    \hline
    -5    & 1.00E-03\footnotemark  & 5.00E-04 & 5.00E-04 & 4.70E-03 & 2.22E-02 & 2.61E-02 \bigstrut\\
    \hline
    -2    & 8.00E-04 & 8.00E-04 & 1.30E-03 & 1.21E-02 & 3.12E-02 & 2.82E-02 \bigstrut\\
    \hline
    1     & 1.00E-04 & 3.00E-04 & 7.00E-04 & 2.13E-02 & 3.75E-02 & 3.04E-02 \bigstrut\\
    \hline
    4     & 1.10E-03 & 2.10E-03 & 9.00E-04 & 3.11E-02 & 4.40E-02 & 3.11E-02 \bigstrut\\
    \hline
    7     & 6.00E-04 & 2.30E-03 & 8.00E-04 & 3.83E-02 & 4.71E-02 & 3.17E-02 \bigstrut\\
    \hline
    10    & 8.00E-04 & 1.60E-03 & 1.20E-03 & 4.47E-02 & 4.81E-02 & 3.15E-02 \bigstrut\\
    \hline
    \end{tabular}%
  \label{tab:addlabel}%
  \begin{tablenotes}
        \footnotesize
        \item[1] The scientific E-notation is used here, for example, 1.00E-03 is equal to $1\times 10^{-3}$.
      \end{tablenotes}
      \end{threeparttable}
\end{table*}%

\begin{figure}[tbp]
\centering
\includegraphics[scale=0.65]{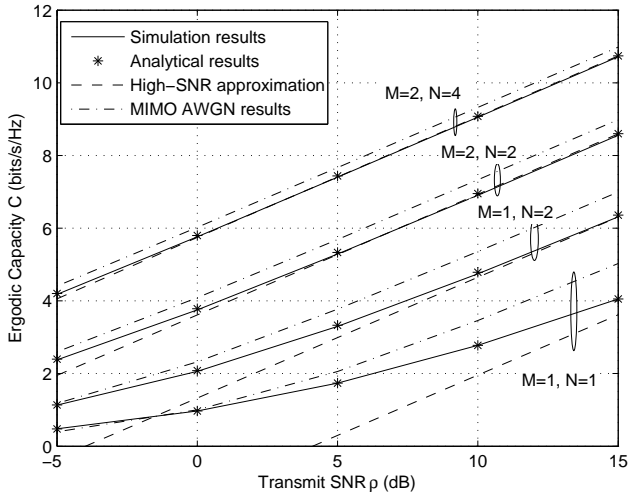}
\caption{Simulated, analytical, and high-SNR approximation ergodic capacity of MIMO FSO systems over i.n.i.d. distributed strong turbulence fading channels versus the transmit SNR ${{\rho }} $ ($C_n^2= 3\times 10^{-14} {\text{m}}^{-2/3}$, $D=0.01$ m, $L=4000$ m, $\lambda=850$ nm and $\eta = 1$). Moreover, the capacity of the non-turbulent channel, e.g., MIMO AWGN channel, is plotted to provide a benchmark for comparison.
\label{fig:Capacity_INID}}
\end{figure}

Figure \ref{fig:Capacity_INID} illustrates the analytical \eqref{eq:inid_capacity_expression}, high-SNR \eqref{eq:inid_capacity_expression_app}, simulated, and AWGN capacity of MIMO FSO systems over i.n.i.d. strong turbulence fading channels. The parameters $k_l$ and $m_l$ is given by (\ref{eq:beta_OC}) and (\ref{eq:m_OC}), respectively, and the expected fading coefficient $\Omega_l$ is randomly chosen subject to a real normal distribution with mean one and standard variance 0.1. As it is clearly illustrated, the analytical results are accurate in the entire SNR regime, indicating the validity of our results. Moreover, the high-SNR expressions become sufficiently tight even at moderate SNRs (e.g., $5$ dB) when deploying only two apertures at both the transmitter and receiver sides. Note that the gap between the AWGN and analytical curves becomes stable as $M$ and $N$ are moderately large.

\begin{figure}[tbp]
\centering
\includegraphics[scale=0.65]{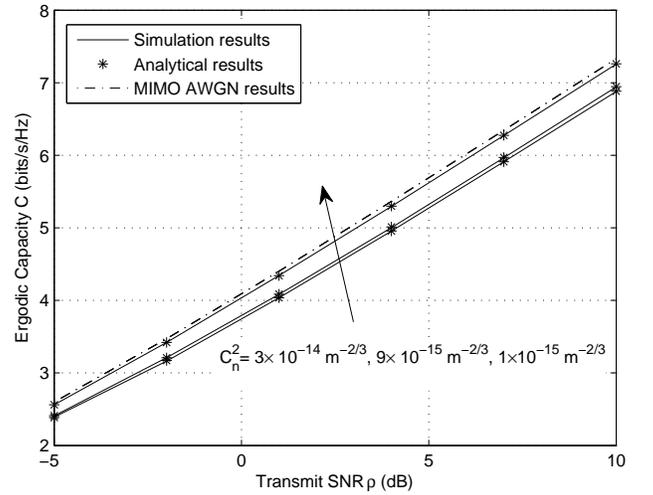}
\caption{Simulated and analytical ergodic capacity of MIMO FSO systems over weak to strong i.i.d. distributed turbulence fading channels versus the transmit SNR ${{\rho }} $ ($M=N=2$, $D=0.01$ m, $L=4000$ m, $\lambda=850$ nm and $\eta = 1$). Moreover, the capacity of the non-turbulent channel, e.g., MIMO AWGN channel, is plotted to provide a benchmark for comparison.
\label{fig:Capacity_C_N}}
\end{figure}

\begin{figure}[tbp]
\centering
\includegraphics[scale=0.65]{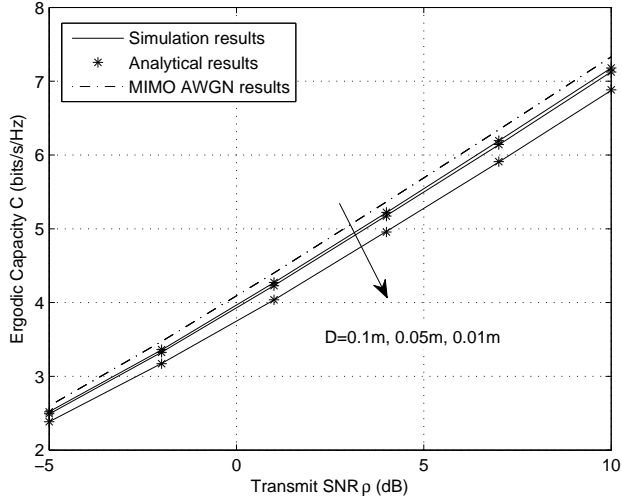}
\caption{Simulated and analytical ergodic capacity of MIMO FSO systems and different receiver aperture diameter $D$ over i.i.d. distributed strong turbulence fading channels versus the transmit SNR ${{\rho }} $ ($M=N=2$, $C_n^2= 3\times 10^{-14} {\text{m}}^{-2/3}$, $L=4000$ m, $\lambda=850$ nm and $\eta = 1$). Moreover, the capacity of the non-turbulent channel, e.g., MIMO AWGN channel, is plotted to provide a benchmark for comparison.
\label{fig:Capacity_D}}
\end{figure}

\begin{figure}[tbp]
\centering
\includegraphics[scale=0.65]{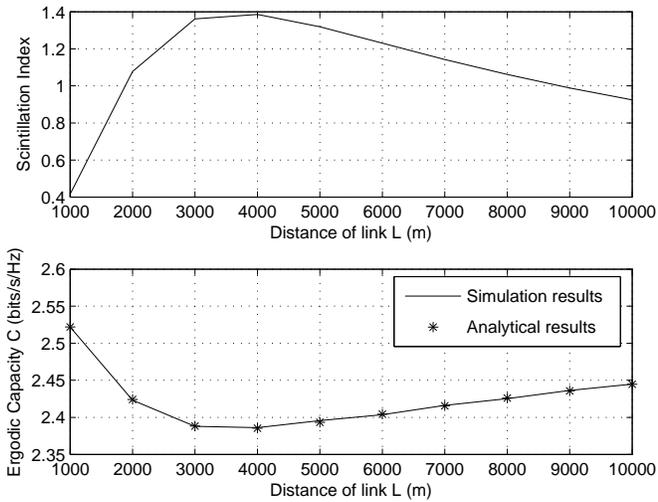}
\caption{Simulated, analytical ergodic capacity and scintillation index of MIMO FSO systems over i.i.d. distributed strong turbulence fading channels versus the distance of link $L$ (${\bar \gamma} =-5$ dB, $M=N=2$, $D=0.01$ m, $C_n^2= 3\times 10^{-14} {\text{m}}^{-2/3}$, $\lambda=850$ nm and $\eta = 1$).
\label{fig:Capacity_L_SI}}
\end{figure}

\begin{figure}[tbp]
\centering
\includegraphics[scale=0.65]{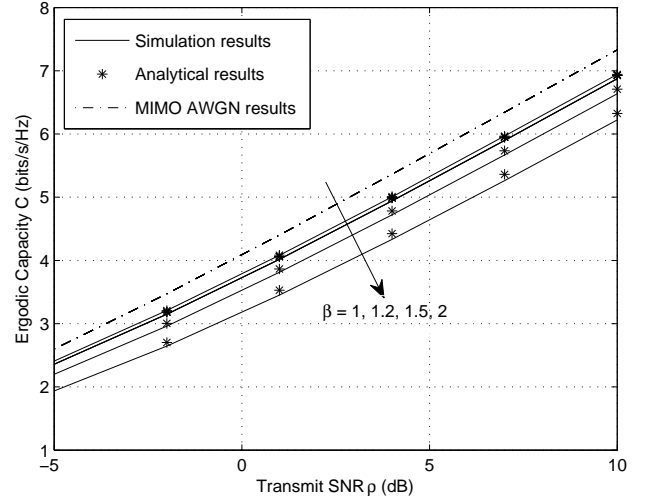}
\caption{Simulated and analytical ergodic capacity of MIMO FSO systems and different fading coefficient ratio $\beta$ over i.n.i.d. distributed strong turbulence fading channels versus the transmit SNR ${{\rho }} $ ($M=N=2$, $C_n^2= 3\times 10^{-14} {\text{m}}^{-2/3}$, $D=0.01$ m, $L=4000$ m, $\lambda=850$ nm and $\eta = 1$). Moreover, the capacity of the non-turbulent channel, e.g., AWGN channel, is plotted to provide a benchmark for comparison.
\label{fig:Capacity_omega}}
\end{figure}

After validating the accuracy of our derived expressions, we then investigate the impact of system parameters on the ergodic capacity. Due to the space limit, we only consider the i.i.d. $\Gamma \Gamma$ fading channels in Figs. \ref{fig:Capacity_C_N}-\ref{fig:Capacity_L_SI}. Note that similar findings can be observed for the i.n.i.d. case. In Fig. \ref{fig:Capacity_C_N}, the ergodic capacity of MIMO FSO systems under i.i.d. $\Gamma \Gamma$ fading conditions is illustrated for various values of the strength of the atmospheric turbulence $C_n^2$. {The arrow in Fig. \ref{fig:Capacity_C_N} means that the corresponding curves denote the value of $C_n^2$ from $3\times 10^{-14} {\text{m}}^{-2/3}$ to $1\times 10^{-15} {\text{m}}^{-2/3}$. The similar method of arrows are used in Figs. \ref{fig:Capacity_D} and \ref{fig:Capacity_omega}.} In all cases, the link distance is fixed as $L=4000$ m with constant $D=0.01$ m and $M=N=2$. The influence of the turbulence strength on the ergodic capacity becomes more obvious from weak to strong turbulence channels. For example, at ${\bar \gamma } = 10$ dB, the difference of the ergodic capacity between $C_n^2= 9\times 10^{-15} {\text{m}}^{-2/3}$ and $C_n^2= 1\times 10^{-15} {\text{m}}^{-2/3}$ is much larger than that between $C_n^2= 3\times 10^{-14} {\text{m}}^{-2/3}$ and $C_n^2= 9\times 10^{-15} {\text{m}}^{-2/3}$. Fig. \ref{fig:Capacity_C_N} indicates that the MIMO FSO systems almost achieve the AWGN capacity under clear weather or weak turbulence conditions (small values of $C_n^2$), while it losses much capacity in adverse weather or strong turbulence conditions (large values of $C_n^2$).

The impact of the receiver aperture diameter $D$ on the ergodic capacity of MIMO FSO systems over i.i.d. distributed strong turbulence fading channels is studied in Fig. \ref{fig:Capacity_D}. In addition, the AWGN capacity is plotted for comparison. The case of $M=N=2$ apertures and $L=4000$ m are considered. The graph likewise indicates that the analytical ergodic capacity coincides with simulated results in all cases under consideration, thereby validating the correctness of the our analytical results. Moreover, the receiver aperture diameter $D$ has an obvious effect on the ergodic capacity, while its impact becomes more obvious as $D$ becomes smaller. This observation can be explained by the fact that larger $D$ tends to drive the fading channels approaching the deterministic case.

In Fig. \ref{fig:Capacity_L_SI}, the effects of optical link distance $L$ on the ergodic capacity and scintillation index of MIMO FSO systems over i.i.d. distributed strong turbulence fading channels are investigated, assuming ${\bar \gamma} =-5$ dB, $M=N=2$, $D=0.01$ m and $C_n^2= 3\times 10^{-14} {\text{m}}^{-2/3}$. It is interesting to see that as $L$ increases, the scintillation index in \eqref{eq:SI} approaches a maximum value greater than $1$, and then changes the slope at a distance of about $4000$ m. We observe that as scintillation index increases, the turbulence effect is getting stronger and thus the ergodic capacity decreases. Indeed, it is not a proportional relationship between the ergodic capacity and link distance as expected. This conclusion is consistent with the results presented in \cite{tsiftsis2009optical,deng2013capacity}. Therefore, the FSO system can achieve a certain capacity gain by adjusting the optical link distance.

Assuming that the received optical signal undergoes i.n.i.d. turbulence channels, and the fading coefficient $\Omega_l$ varies in each link. We consider a typical case of $\Omega_l=\beta\Omega_{l-1}\;(2\le l\le L)$, where $\beta\ge1$ is the ratio of adjacent fading coefficients, to investigate the effect of fading coefficient volatility on the ergodic capacity. {For the considered MIMO FSO system with AWGN receiver noise and EGC receiver, the relationship between the ergodic capacity and $\beta$ is studied in Fig. \ref{fig:Capacity_omega}}, where the analytical results are generated from \eqref{eq:capacity_egc_result_H} for i.i.d. $\Gamma\Gamma$ channels (e.g., $\beta=1$) and \eqref{eq:inid_capacity_expression} for i.n.i.d. $\Gamma\Gamma$ channels (e.g., $\beta \ge 1$). For easy tractability but without loss of generality, all $\Omega_l$s are normalized to yield an identical electrical SNR given by (\ref{eq:EGC_SNR}). From Fig. \ref{fig:Capacity_omega}, it is clear that the ergodic capacity is descending as $\beta$ increases {(e.g., the ergodic capacity for $\beta=2$ is much lower than that for $\beta=1$)}, or equivalently, the ergodic capacity is shown to be larger when $\Omega_l$ from different links are more balanced, i.e., with lower volatility. Moreover, there is a significant reduction in ergodic capacity when one of the receive apertures collect most of the transmission power, such as $\beta=2$ where $\Omega_L>\sum_{l=1}^{L-1}\Omega_l$. This observation can be easily interpreted due to the EGC scheme used in our analysis. {Finally, it is worthy to note from Fig. \ref{fig:Capacity_omega} that the analytical results are not very accurate with $\beta=1.5$ and $\beta=2$. However, this approximation acts as a lower bound for all cases.}

\section{Conclusions}\label{se:conclusion}
{In this paper, the ergodic capacity performance of MIMO FSO systems with the EGC scheme turbulence channels was studied. An $\alpha$-$\mu$ approximation has been used for the sum of i.i.d. $\Gamma\Gamma$ RVs, while a nested finite weighted approximation was employed for the sum of i.n.i.d. $\Gamma\Gamma$ RVs. Analytical and accurate expressions of the ergodic capacity in forms of Meijer's $G$- and Fox's $H$-functions were derived for both i.i.d. and i.n.i.d. cases. Accurate asymptotic ergodic capacity expressions at the high-SNR regime were also derived to provide useful insights regarding the parameters that affect the system performance. For example, a capacity gain can be achieved by using more transmit and receive apertures as well as increasing the diameter of receive apertures. Interestingly, for practical MIMO FSO systems, it is not a proportional relationship between the ergodic capacity and link distance as expected. Moreover, analytical results accompanied with Monte-Carlo simulations were presented to demonstrate effectiveness of the proposed expressions. Finally, the analysis method proposed in this paper can be used to derive the ergodic capacity of MIMO FSO system for other receiver scheme, such as MRC.}


\bibliographystyle{IEEEtran}
\bibliography{IEEEabrv,Ref}

\end{document}